\documentclass[aps,pre,preprint,superscriptaddress]{revtex4-1}  
\usepackage{graphicx}
\usepackage{dcolumn}
\usepackage{amssymb}
\usepackage{amsmath}
\usepackage{newtxtext,newtxmath,lipsum}
\usepackage[shortlabels]{enumitem}

\newcommand{\sech}{\normalfont\mbox{sech}\,}

\begin{document}
	\title{On examining the predictive capabilities of two variants of PINN in validating localised wave solutions in the generalized nonlinear Schr\"{o}dinger equation}
	\author{K. Thulasidharan} 
	\affiliation{Department of Physics, Vellore Institute of Technology, Vellore - 632 014, Tamilnadu, India}
	
	\author{N. Sinthuja}
	\affiliation{Department of Physics, Anna University, Chennai - 600 025, Tamilnadu, India}
	
	\author{N. Vishnu Priya}
	\affiliation{Department of Mathematics, Indian Institute of Science, Bangalore - 560012, Karnataka, India}
	
	\author{M. Senthilvelan}
	\affiliation{Department of Nonlinear Dynamics, Bharathidasan University, Tiruchirappalli - 620 024, Tamilnadu, India}
	\email{velan@cnld.bdu.ac.in}
	\begin{abstract}
		\par  We introduce a novel neural network structure called Strongly Constrained Theory-Guided Neural Network (SCTgNN), to investigate the behaviours of the localized solutions of the generalized nonlinear Schr\"{o}dinger (NLS) equation. This equation comprises four physically significant nonlinear evolution equations, namely, (i) NLS equation, Hirota equation Lakshmanan-Porsezian-Daniel (LPD) equation and fifth-order NLS equation. The generalized NLS equation demonstrates nonlinear effects up to quintic order, indicating rich and complex dynamics in various fields of physics. By combining concepts from the Physics-Informed Neural Network (PINN) and Theory-Guided Neural Network (TgNN) models, SCTgNN aims to enhance our understanding of complex phenomena, particularly within nonlinear systems that defy conventional patterns. To begin, we employ the TgNN method to predict the behaviours of localized waves, including solitons, rogue waves, and breathers, within the generalized NLS equation. We then use SCTgNN to predict the aforementioned localized solutions and calculate the mean square errors in both SCTgNN and TgNN in predicting these three localized solutions. Our findings reveal that both models excel in understanding complex behaviours and provide predictions across a wide variety of situations.
	\end{abstract}
	
	\maketitle
	\section{Introduction}
The one-dimensional fundamental nonlinear Schr\"{o}dinger (NLS) equation is renowned for its integrability, a property that has catalyzed significant advancements in the analysis of nonlinear phenomena across various disciplines, including optics, water waves, and Bose-Einstein condensates \cite{1,2,3}. The availability of analytical solutions for the NLS equation has not only fueled theoretical developments but has also inspired numerous experimental investigations in these domains. While soliton solutions marked initial milestones, recent progress has been marked by advancements in breather and rogue wave solutions, showcasing the versatility of the NLS equation \cite{4}.

However, the NLS equation stands as just one among several integrable nonlinear evolutionary equations in physics. Various extensions and deformations of the NLS equation have been explored, broadening its scope of application. These extensions not only enhance our understanding of fundamental physics but also provide insights into complex phenomena like wave blow-up and collapse at higher intensities \cite{5}, necessitating the inclusion of higher-order terms \cite{6,7,8}. On other hand, in optics, in the femtosecond range, the third-order effect becomes vital \cite{9}, while the fourth-order effect's role in an anisotropic Heisenberg ferromagnetic spin has been examined in \cite{10,11,12}. Notably, the quintic effect gains significance as the optical field's intensity escalates and pulse duration contract to attosecond scales \cite{13,14}.  The generalized NLS equation introduced in \cite{15} included all these higher order effects.  

In this study, we consider the NLS equation to fifth-order terms, constituting the NLSE hierarchy. Our approach involves augmenting the NLSE with additional terms possessing arbitrarily large coefficients. This extension results in a hierarchy of integrable equations up to the fifth order, each characterized by a set of real coefficients. The supplementary terms encompass higher-order dispersion of all orders and nonlinear terms, surpassing the simplicity of the NLSE. The flexibility conferred by arbitrary coefficients enables us to explore nonlinear phenomena with unprecedented depth, shedding light on intricate physical processes beyond the scope of conventional NLSE formulations. The generalized NLS equation also admits several kinds of localized solutions - solitons, rogue waves, and breathers, to name a few \cite{15}. These nonlinear waves categorize a wide range of diverse physical fields, including fluid dynamics, optics, and plasma physics, making the generalized NLS equation a versatile tool for understanding and describing complex wave phenomena.

\par Solitons were first identified in the context of water waves and now they are found in almost all scientific fields.  These waves maintain same speed and shape while propagating.  Solitons possess numerous applications in the real world, see for instance \cite{16}.  Rogue waves, originally coined to describe extreme ocean events, has drawn considerable interest through both experimental observations and theoretical predictions. Such events manifest in nonlinear fiber optics, Bose-Einstein condensations (BECs), plasmas, and even financial contexts \cite{17,18,19,20}. These waves, with amplitudes often exceeding double the significant wave height, appear suddenly and vanish without a trace. Although their origin remains unclear, a consensus among most researchers suggests that these entities are associated with specific types of waves produced by mathematical equations. These special waves, known as `breathers', could be early signs of rogue waves because they develop when small disturbances grow into significant ones. Generally, two breather structures exist: the Akhmediev breather (AB) and the Kuznetsov-Ma breather (KMB) \cite{15,21}. The AB represents a space - periodic wave localized in time, while the KMB is localized in space and oscillates periodically in time. In specific conditions, both structures become the rogue wave solution of the NLS equation. These breathers serve as plausible models for comprehending the dynamics of rogue waves across diverse physical realms. In the literature, multiple methods have been employed, say for example, inverse scattering transform \cite{22,23}, Hirota bilinearization \cite{24}, and Darboux transformation \cite{21,25} in order to obtain these solutions.

In recent years, artificial intelligence (AI) and machine learning (ML) have gained widespread application in efficiently managing large data sets and have assumed increasingly significant roles across various domains \cite{26,27}. A recent development involves employing deep neural networks to explore data-oriented solutions and identify parameters within nonlinear physical models, including the fractional version of nonlinear systems \cite{28,29,30,31,32,33,34,35}. The concept of physics-informed neural networks (PINNs) has emerged as a technique for investigating nonlinear partial differential equations \cite{36,37}. Utilizing the deep learning method PINN one can attain precise solutions with minimal data. Simultaneously, since fundamental physical constraints are typically expressed via differential equations, this approach also offers a more comprehensive physical rationale for the predicted solution \cite{38,39,40}. 
Strongly-Constrained Physics - Informed Neural Network (SCPINN) \cite{41,42}, Adaptive PINN \cite{43,44} and Theory-Guided Neural Network (TgNN) \cite{45,46} are variants of PINN. In the following, we point out the differences between these three network structures.\\
\textbf{PINN:} The PINN approach integrates established physical equations or constraints into the neural network's learning process. This fusion of data-driven learning with domain-specific principles enhances predictions. It proves particularly useful for resolving challenges arising from limited or noisy data, as well as scenarios where the underlying physics is less understood. Its applications span diverse fields, enabling prediction of physical behaviours, system simulations, and solving differential equations \cite{36,37}. \\
\textbf{SCPINN:} SCPINN advances the PINN approach by employing more robust and stringent constraints based on derivative information from solutions. It introduces parallel subnets, adaptive weights, and flexible learning rates to overcome limitations present in standard PINN. The result is a remarkable improvement in prediction accuracy across broader computational domains, rendering it well-suited for intricate problems \cite{41,42}.\\
\textbf{TgNN:} TgNN encapsulates a broader notion, wherein neural networks are guided by domain-specific theories or principles, extending beyond physics. It incorporates established theories or constraints to amplify predictions beyond the scope of data-driven learning alone. Its application is wide-ranging, spanning domains such as economics, biology, engineering, and more. This approach amalgamates existing knowledge with machine learning, fostering precise and insightful predictions \cite{45,46}.

In summary, PINN combines physics and neural networks for smart predictions. SCPINN makes this mix even stronger. TgNN is a broader approach that covers various smart networks, not just for physics. The methods we mentioned earlier, TgNN and SCPINN, are useful tools that help one to make computer predictions better in different areas. They do this by combining data-based learning with expert knowledge, leading to important improvements.

In our current study, we intend to combine the strong points of SCPINN and TgNN to make a new method called Strongly Constrained Theory Guided Neural Network (SCTgNN). This new type of neural network help us to understand complicated structures like rogue waves and breathers in more detail, which in turn will help us to understand nonlinear systems more deeply. In \cite{45}, the authors used the TgNN model to predict rogue wave solutions. This TgNN model was more accurate compared to the convolution neural network (CNN) model. But the authors in \cite{45} have studied only the prediction of rogue waves in the basic NLS equation. In contrast, our goal is to go beyond that and explore solitons, rogue waves, and breathers for a more general equation (generalized NLS equation). 

Through our investigation, we affirm that the incorporation of higher-order dispersion parameters into the standard NLS equation induces noteworthy alterations in the dimensions and alignment of solitons, rogue waves, and breathers. Intriguingly, both the SCTgNN and TgNN models consistently demonstrate precise error predictions. By integrating system parameters $\alpha_2$, $\alpha_3$, $\alpha_4$, and $\alpha_5$ into the initial conditions, we attain a significant advantage on the direct forecasting of solutions across various parameter values.

These findings underscore the effectiveness of the novel SCTgNN model in accurately approximating soliton, rogue waves, and breather solutions within the broader generalized NLS equation. This holds true even under more intricate scenarios involving higher-order effects, showcasing the versatility and robustness of our approach.

The structure of our manuscript is outlined as follows: In Section 2, we illustrate the TgNN and SCTgNN models, along with the methods employed. Section 3 is dedicated to showcasing the data-driven soliton, rogue wave, and breather solutions for the generalized NLS equation and the predictions made. Our findings are summarized in Section 4.

\section{Model and method}
We consider the generalized nonlinear Schr\"odinger equation in the form \cite{15,47}
\begin{subequations}
	\label{m1am2}
	\begin{align}
	i\psi_x+\alpha_2\Gamma_2[\psi(x,t)]-i\alpha_3\Gamma_3[\psi(x,t)]+\alpha_4\Gamma_4[\psi(x,t)]-i\alpha_5\Gamma_5[\psi(x,t)]=0,
	\label{m1}
	\end{align}
	with
	\begin{align}
	\Gamma_2[\psi(x,t)]=&\psi_{tt}+2|\psi|^2\psi,\\
	\Gamma_3[\psi(x,t)]=&\psi_{ttt}+6|\psi|^2\psi_t,\\
	\Gamma_4[\psi(x,t)]=&\psi_{tttt}+8|\psi|^2\psi_{tt}+6|\psi|^4\psi+4\psi|\psi_t|^2+6\psi^2_t\bar{\psi}+2\bar{\psi}_{tt}\psi^2,\\
	\Gamma_5[\psi(x,t)]=&\psi_{ttttt}+10|\psi|^2\psi_{ttt}+30|\psi|^4\psi_t+10\psi\psi_t\bar{\psi}_{tt}+10\psi\bar{\psi}_t\psi_{tt}+20\bar{\psi}\psi_t\psi_{tt}\nonumber\\&+10\psi^2_t\bar{\psi}_t.
	\label{m2}
	\end{align}
\end{subequations}
In Eq. (\ref{m1}), $x$ and $t$ are the propagation and transverse variable respectively. The function $\psi(x,t)$ denotes the envelope of the waves. The coefficients $\alpha_i$ ($i=2,3,4,5$) are arbitrary real constants. This generalized form of equation (\ref{m1}) augments four completely integrable systems separately. If we consider the first term and the lowest second order terms $\Gamma_2[\psi(x,t)]$, we can obtain the fundamental NLS equation (the other parameters are $\alpha_3=0, \alpha_4=0, \alpha_5=0$). The first term with $\Gamma_3[\psi(x,t)]$, gives the Hirota equation ($ \alpha_4=0, \alpha_5=0$) and if we choose the first term with $\Gamma_4[\psi(x,t)]$, we get the fourth order NLS equation ($ \alpha_3=0, \alpha_5=0$). Similarly with $\Gamma_5[\psi(x,t)]$, we arrive at a fifth-order NLS equation ($\alpha_3=0, \alpha_4=0$).
\begin{figure*}[!ht]
	\begin{center}
		\includegraphics[width=\linewidth]{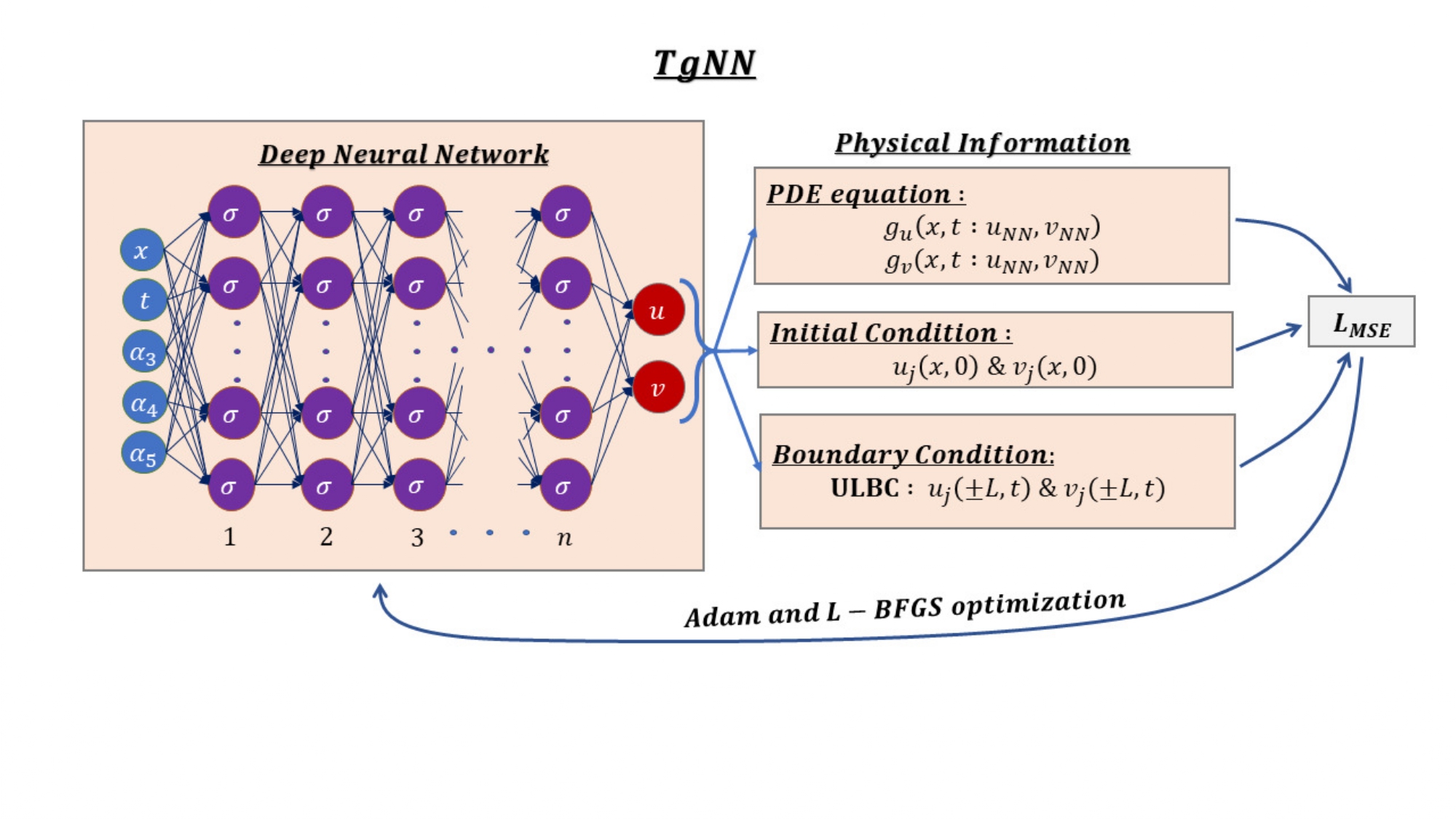}
	\end{center}
	\vspace{-1cm}
	\caption{A graphical representation of the TgNN showcases its intricate structure, which comprises of input layer, hidden layers, and an output layer. Neurons are represented as circles, and the connections between them are depicted with arrows, signifying the functional mappings. The input variables include $x$, $t$, $\alpha_3$, $\alpha_4$ and $\alpha_5$ while the output layer provides $u$ and $v$, which correspond to the real and imaginary components of the wave envelope $\psi(x,t)$. While the deep learning process of TgNN is data-driven, it is equally influenced by the generalized NLS equation (PDE), along with the initial condition and boundary condition.}
	\label{fig01}
\end{figure*}
\begin{figure*}[!ht]
	\begin{center}
		\includegraphics[width=0.7\linewidth]{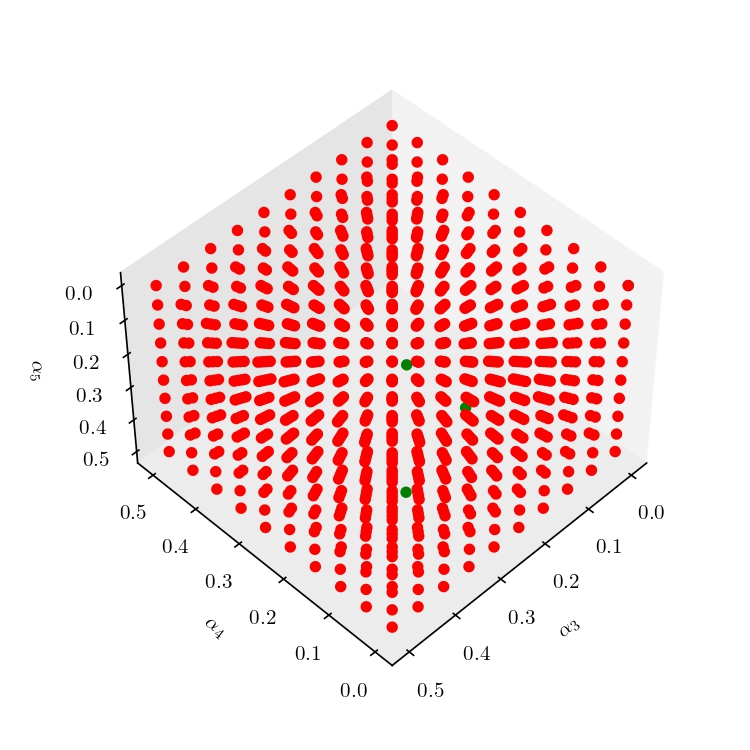}
	\end{center}
	\vspace{-0.3cm}
	\caption{The red dots on the graph symbolize the dataset obtained through the traditional real-time evolution method for training purposes. To assess and compare the proficiency of the successfully trained neural network, we employ the green dots, which correspond to the situations depicted in Figures \ref{fig2}, \ref{fig3} and \ref{fig4}.}
	\label{fig03}
\end{figure*} 

The fifth-order NLS equation has been the subject of extensive research over the years. Chowdury et al. presented the fifth-order NLS equation along with its Lax pair and constructed soliton solutions using the Darboux transformation method \cite{48}. This equation is widely recognized as a model describing the one-dimensional anisotropic Heisenberg ferromagnetic spin chain \cite{49}. Subsequently, breathers and rogue wave solutions for the fifth-order NLS equation were obtained through the Darboux transformation \cite{6}. The fifth-order equation also finds applications in describing the propagation of ultrashort optical pulses in optical fibers \cite{7}. Wang discussed the structures of higher-order rogue wave solutions and their interactions \cite{7}. Several studies have focused on deriving analytic solutions for the fifth-order NLS equation. Infinitely-many conservation laws were established based on the Lax pair, and analytic solutions, including one-, two-, and three-soliton forms, were also obtained using auxiliary functions method, the Hirota bilinear method, and symbolic calculation methods \cite{8}. Yang et al. provided first- and second-order rogue-wave solutions, as well as rational solitons \cite{50}, while dark one-, two-, and three-soliton solutions were generated using the Hirota bilinear method \cite{51}. The generalized Darboux transformation was utilized to derive first-, second-, and third-order rogue wave solutions \cite{52}. Jia investigated the interaction and propagation of three kinks of breather solutions and analyzed the modulation instability of generalized solitons \cite{53}. Subsequently, various N-soliton solutions, bright N-soliton solutions, N-dark soliton solutions, and higher-order rogue wave solutions were successively obtained \cite{54}. Sinthuja et al. studied the formation of rogue waves on a periodic background in the fifth-order NLS equation\cite{13}. In the following, we move on to analyze how the generalized fifth-order NLS equation is modeled by TgNN and SCTgNN.

As a first step, we split the system into its real and imaginary components. This involves separating the complex wave envelope, denoted as $\psi(x,t)$, into two real functions: $u(x,t)$ and $v(x,t)$. This separation can be achieved by representing $\psi(x,t)$ in the form $\psi(x,t) = u(x,t) + iv(x,t)$, where $u(x,t)$ and $v(x,t)$ are the real components.

Once this separation is established, we can proceed with detailing the structure of the machine learning model. However, to set the stage, it is essential to start by substituting the aforementioned form of $\psi(x,t)$ into the equation of interest, Eq. (\ref{m1am2}). This substitution allows us to explicitly derive the expressions for $u(x,t)$ and $v(x,t)$, which take the following forms:
\begin{subequations}
	\label{m3am4}
	\begin{align}
	v_x=&~\alpha_2(u_{tt}+2u^3+2uv^2)+\alpha_3(v_{ttt}+6u^2v_t+6v^2v_t)+\alpha_4(6u^5+12u^3v^2+6uv^4\nonumber\\&+10uu^2_t+12vu_tv_t-2uv^2_t+10u^2u_{tt}+6v^2u_{tt}+4uvv_{tt}+u_{tttt})+\alpha_5(30u^4v_t+60\nonumber\\&\times u^2v^2v_t+30v^4v_t+10u^2_tv_t+10v^3_t+20uv_tu_{tt}+20uv_{tt}u_t+40v_tvv_{tt}+10u^2v_{ttt}\nonumber\\&+10v^2v_{ttt}+v_{ttttt}),
	\label{m3}\\
	-u_x=&~\alpha_2(v_{tt}+2u^2v+2v^3)+\alpha_3(-6u^2u_t-6v^2u_t-u_{ttt})+\alpha_4(6u^4v+12u^2v^3+6v^5\nonumber\\&-2vu^2_t+12uu_tv_t+10vv_t+4uvu_{tt}+6u^2v_{tt}+10v^2v_{tt}+v_{tttt})+\alpha_5(-30u^4u_t-60\nonumber\\&\times u^2v^2u_t-30v^4u_t-10u^3_t-10u_tv^2_t-40uu_tu_{tt}-20vv_tu_{tt}-20vu_tv_{tt}-10u^2u_{ttt}\nonumber\\&-10v^2u_{ttt}-u_{ttttt}).
	\label{m4}
	\end{align}
\end{subequations}
and the solution $\psi(x,t)$ is trained to satisfy the neural networks  (\ref{m3}) and (\ref{m4}).
\begin{figure*}[!ht]
	\begin{center}
		\includegraphics[width=\linewidth]{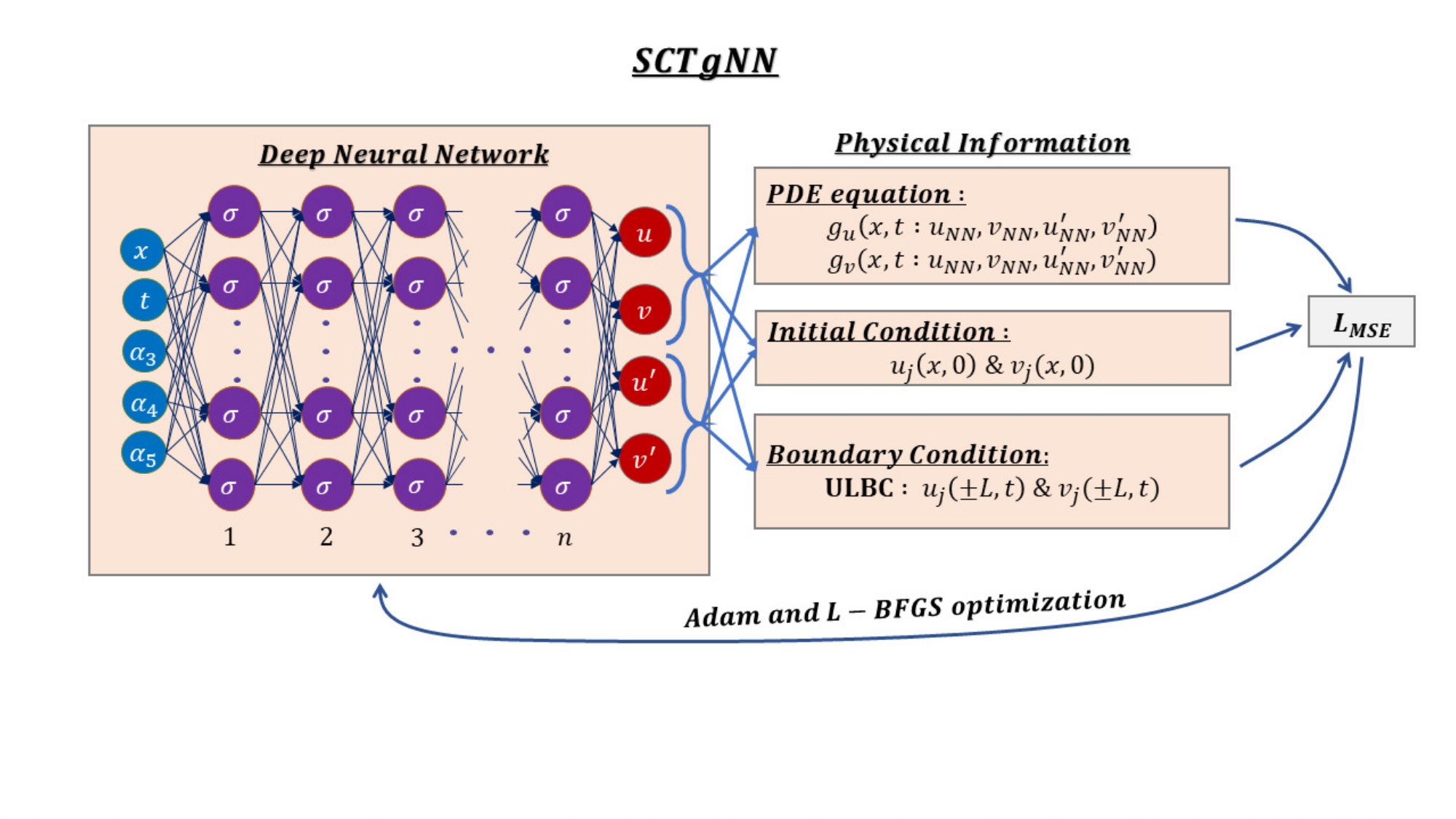}
	\end{center}
	\vspace{-1cm}
	\caption{The graphical representation of the SCTgNN showcases its intricate architecture, consisting of an input layer, hidden layers, and an output layer. While the SCTgNN model shares similarities with TgNN, it differs in that TgNN's output layer yields both u and v, whereas SCTgNN goes beyond and produces $u^{\prime}$ and $v^{\prime}$.}
	\label{fig02}
\end{figure*} 

The above equation (\ref{m3am4}) can be rewritten in the following form
\begin{subequations}
	\begin{align}
	f_u(x,t)=&-v^{NN}_x(.)+\alpha_2[u^{NN}_{tt}(.)+2u{^{3}}^{NN}(.)+2u^{NN}(.)v{^2}^{NN}(.)]+\alpha_3[v^{NN}_{ttt}(.)+6u{^2}^{NN}(.)\nonumber\\&\times v^{NN}_t(.)+6v{^2}^{NN}(.)v^{NN}_t(.)]+\alpha_4[6u{^5}^{NN}(.)+12u{^3}^{NN}(.)v{^2}^{NN}(.)+6u^{NN}(.)\nonumber\\&\times v{^4}^{NN}(.)+10u^{NN}(.) u{^2}^{NN}_t(.)+12v^{NN}(.)u^{NN}_t(.)v^{NN}_t(.)-2u^{NN}(.)v{^2}^{NN}_t(.)\nonumber\\&+10u{^2}^{NN}(.)u^{NN}_{tt}(.)+6v{^2}^{NN}(.) u^{NN}_{tt}(.)+4u^{NN}(.)v^{NN}(.)v^{NN}_{tt}(.)+u^{NN}_{tttt}(.)]\nonumber\\&+\alpha_5[30u{^4}^{NN}(.)v^{NN}_t(.)+60u{^2}^{NN}(.)v{^2}^{NN}(.) v^{NN}_t(.)+30v{^4}^{NN}(.)v^{NN}_t(.)\nonumber\\&+10u{^2}^{NN}_t(.)v^{NN}_t(.)+10v{^3}^{NN}_t(.)+20u^{NN}(.)v^{NN}_t(.)u^{NN}_{tt}(.)+20u^{NN}(.)v^{NN}_{tt}(.)\nonumber\\&\times u^{NN}_t(.)+40v^{NN}_t(.)v^{NN}(.)v^{NN}_{tt}(.)+10u{^2}^{NN}(.)v^{NN}_{ttt}(.)+10v{^2}^{NN}(.) v^{NN}_{ttt}(.)\nonumber\\&+v^{NN}_{ttttt}(.)],
	\label{m33}
	\end{align}
	\begin{align}
	f_v(x,t)=&~u^{NN}_x(.)+\alpha_2[v^{NN}_{tt}(.)+2u{^2}^{NN}(.)v^{NN}(.)+2v{^3}^{NN}(.)]+\alpha_3[-6u{^2}^{NN}(.)u^{NN}_t(.)\nonumber\\&-6v{^2}^{NN}(.) u^{NN}_t(.)-u^{NN}_{ttt}(.)]+\alpha_4[6u{^4}^{NN}(.)v^{NN}(.)+12u{^2}^{NN}(.)v{^3}^{NN}(.)\nonumber\\&+6v{^5}^{NN}(.)-2v^{NN}(.) u{^2}^{NN}_t(.)+12u^{NN}(.)u^{NN}_t(.)v^{NN}_t(.)+10v^{NN}(.)v^{NN}_t(.)\nonumber\\&+4u^{NN}(.)v^{NN}(.)u^{NN}_{tt}(.)+6u{^2}^{NN}(.) v^{NN}_{tt}(.)+10v{^2}^{NN}(.)v^{NN}_{tt}(.)+v^{NN}_{tttt}(.)]\nonumber\\&+\alpha_5[-30u{^4}^{NN}(.)u^{NN}_t(.)-60u{^2}^{NN}(.)v{^2}^{NN}(.) u^{NN}_t(.)-30v{^4}^{NN}(.)u^{NN}_t(.)\nonumber\\&-10u{^3}^{NN}_t(.)-10u^{NN}_t(.)v{^2}^{NN}_t(.)-40u^{NN}(.)u^{NN}_t(.)u^{NN}_{tt}(.)-20v^{NN}(.)\nonumber\\&\times v^{NN}_t(.)u^{NN}_{tt}(.)-20v^{NN}(.)u^{NN}_t(.)v^{NN}_{tt}(.)-10u{^2}^{NN}(.)u^{NN}_{ttt}(.)-10v{^2}^{NN}(.)\nonumber\\&\times u^{NN}_{ttt}(.)-u^{NN}_{ttttt}(.)],
	\label{m44}
	\end{align}
\end{subequations}
where $^{NN}(.)$ denotes the $NN$ approximation of the real $(u)$ and imaginary $(v)$ parts of the solution $\psi(x,t)$. Here, the partial derivatives can be easily computed by applying the chain rule for the network through automatic differentiation.

Figure \ref{fig01} provides a detailed description of the structure of TgNN. To predict the solution of the generalized NLS equation using TgNN, we employ the hyperbolic tangent function (tanh) as an activation function, with $6$ hidden layers, each comprising $60$ neurons. In TgNN, we incorporate system parameters as inputs to the network. This approach allows us to predict solutions for various values of the system parameters. The system parameters of the generalized NLS equation, such as $\alpha_3$, $\alpha_4$ and $\alpha_5$, are included in the input layer. Consequently, we have a total of five input neurons. The system's domain is defined as follows: $x$ ranges from $-10$ to $10$, $t$ ranges from $-10$ to $10$, $\alpha_3$ varies from $0.0$ to $0.5$, $\alpha_4$ ranges from $0.0$ to $0.5$, and $\alpha_5$ spans from $0.0$ to $0.5$. For training data, we consider $1000$ combinations of system parameters. From these combinations, we extract 2000 sample points from the initial value at $t=0$, as well as $2000$ sample points from the boundaries at $x=-10$ and $x=10$. Additionally, our sample dataset comprises 10000 collocation points that cover the entire domain ($x$, $t$, $\alpha_3$, $\alpha_4$, $\alpha_5$) and enforce the constraint stated in Eq. (\ref{m3am4}). The central coordinates of these regions are determined using a Latin Hypercube Sampling (LHS) strategy \cite{41}.

In Fig. \ref{fig03}, we display the 1000 combinations of training system parameters. The red dots represent the training data points, while the green dots represent the predicted values. In this paper, we present a limited number of predicted plots; however, once the model is trained, it possesses the capability to predict solutions for different parameter combinations. The output layer of the model comprises two neurons that yield the real $(u)$ and imaginary $(v)$ parts of the solution. These solutions undergo automatic differentiation and are then substituted into the loss function. For optimization during back-propagation, we employ both the Adam and L-BFGS methods. The loss function of TgNN can be expressed in the following form:
\begin{subequations}
	\label{l1a}
	\begin{align}
	Loss_{TgNN}=Loss_{I}+Loss_{B}+Loss_{D},
	\label{l1}
	\end{align}
	where
	\begin{align}
	Loss_I=&\frac{1}{N_I}\sum_{j=1}^{N_I}\left(|u^{NN}(x_I^j,0,\alpha_3^j,\alpha_4^j,\alpha_5^j)-u_0^j|^2+|v^{NN}(x_I^j,0,\alpha_3^j,\alpha_4^j,\alpha_5^j)-v_0^j|^2\right),\label{l2}\\
	Loss_B=&\frac{1}{N_B}\sum_{j=1}^{N_B}\left(|u^{NN}(\pm 10,t_B^j,\alpha_3^j,\alpha_4^j,\alpha_5^j)-u_B^j|^2+|v^{NN}(\pm 10,t_B^j,\alpha_3^j,\alpha_4^j,\alpha_5^j)-v_B^j|^2\right),\label{l3}\\
	Loss_{D}=&\frac{1}{N_D}\sum_{j=1}^{N_D}\left(|f_u(x_D^j,t_D^j,\alpha_3^j,\alpha_4^j,\alpha_5^j)|^2+|f_v(x_D^j,t_D^j,\alpha_3^j,\alpha_4^j,\alpha_5^j)|^2\right).\label{l4}
	\end{align}
\end{subequations}

In the above equation, $N_I$ represents the number of collocation points taken in the initial region, $N_B$ represents the number of collocation points taken at the boundaries, and $N_D$ represents the number of collocation points taken in the domain. Additionally, $u^{NN}(x^j, t^j)$ and $v^{NN}(x^j, t^j)$ are the outputs of the model, while $u_0^j$ and $v_0^j$ represent the exact values of the initial frame, respectively. In the second term of Eq. (\ref{l3}), the notations $u_B$ and $v_B$ pertain to the boundary region. In the same way, the collocation points (domain values) for $f_u(x,t)$ and $f_v(x,t)$ are denoted as $\{x^i_{D},t^i_{D}\}$. These collocation points are sampled with the help of the classical LHS technique. 

In Eq. (\ref{l1a}), the loss function is based on the initial and boundary value data as well as the residuals gathered from Eqs. (\ref{m3}) and (\ref{m4}) at a finite set of collocation points sampled. Particularly, in the right hand side of Eq. (\ref{l1}), the first two terms play a role on fitting the solution data, while the other term play a major role to satisfy the residuals $f_u$ and $f_v$.

To compare with TgNN, we adopt the same activation function, number of hidden layers, and neurons in the SCTgNN model which is shown in Fig. \ref{fig02}. The advantage of the SCTgNN model lies in its ability to predict the solution's derivatives. Differing from TgNN, the SCTgNN model has four outputs that provide $u$, $u^{\prime}$, $v$, and $v^{\prime}$ values, where the prime denotes differentiation with respect to $t$. By utilizing a mean squared loss function, the loss of the SCTgNN model can be formulated as follows:
\begin{subequations}
	\label{m5am6}
	\begin{align}
	Loss_{SCTgNN}=Loss_{I}+Loss_{B}+Loss_{D},
	\label{m5}
	\end{align}
	where
	\begin{align}
	\begin{split}
	Loss_{D}=\frac{1}{N_D}\sum_{i=1}^{N_D}\left(|f_u(x_D^i,t_D^i,\alpha_3^i,\alpha_4^i,\alpha_5^i;u,v)|^2+|f_v(x_D^i,t_D^i,\alpha_3^i,\alpha_4^i,\alpha_5^i;u,v)|^2\right. \\
	\left.+|f_{u^{\prime}}(u,u^{\prime},u_t)|^2+|f_{v^{\prime}}(v, v^{\prime},v_t)|^2\right),\label{m6}
	\end{split}
	\end{align}
	in which 
	\begin{align}
	f_{{u}^{\prime}}={u^{\prime}}^{NN}(.)-{u_t}^{NN}(.),\quad f_{v^{\prime}}={v^{\prime}}^{NN}(.)-{v_t}^{NN}(.)\label{m5andm6}.
	\end{align}
\end{subequations}

In equation (\ref{m5am6}), $u^{\prime}$ and $v^{\prime}$ correspond to the third and fourth output of the model, whereas $u_t$ and $v_t$ represent the differentiation with respect to $t$ for the first and second outputs of the model. By incorporating these outputs, the introduced model achieves enhanced optimization compared to TgNN. In SCTgNN, the first two loss terms are similar to those in TgNN, contributing to fitting the solution data. Meanwhile, the last term plays a crucial role in satisfying the residuals $f_u$, $f_v$, ${f_u^{\prime}}$, and ${f_v^{\prime}}$.

In simple terms, both the models can make predictions. SCTgNN stands out for its ability to predict derivatives. The loss function of these models help to improve predictions and meet requirements, ensuring accurate results.

\section{Data driven soliton, rogue wave and breather solutions for the generalized NLS equation}
In this section, we explain three intresting localized solutions, namely solitons, rogue waves, and breathers. We also show the results obtained through SCTgNN and TgNN models, along with the exact solution, for the system described by Eq. (\ref{m1am2}).
\subsection{Soliton solution}
The first-order soliton solution for the equation (\ref{m1am2}) is given by \cite{15}
\begin{align}
\psi_{s}=c\; \sech(xv_s+ct)e^{ix\phi_s},
\label{m10}
\end{align}
where the phase ($v_s$) and velocity ($\phi_s$) having the following form
\begin{align}
v_s=c^3(\alpha_3+c^2\alpha_5),\quad	\phi_s=c^2(\alpha_2+c^2\alpha_4).
\label{m11}
\end{align}
The parameter $c$ is an arbitrary constant. It is noted that the velocity $(v_s)$ described by the third- and fifth-order coefficients, while the phase $(\phi_s)$ is given interms of second- and fourth-order coefficients.

If we consider $t=0$, the same solution (\ref{m10}) turns out to be                                                            
\begin{align}
\psi_{s}=c\; \sech(xv_s)e^{ix\phi_s},
\label{m13}
\end{align}
where $v_s$ and $\phi_s$ are given in Eq. (\ref{m11}).
\begin{figure*}[!ht]
	\begin{center}
		\includegraphics[width=\linewidth]{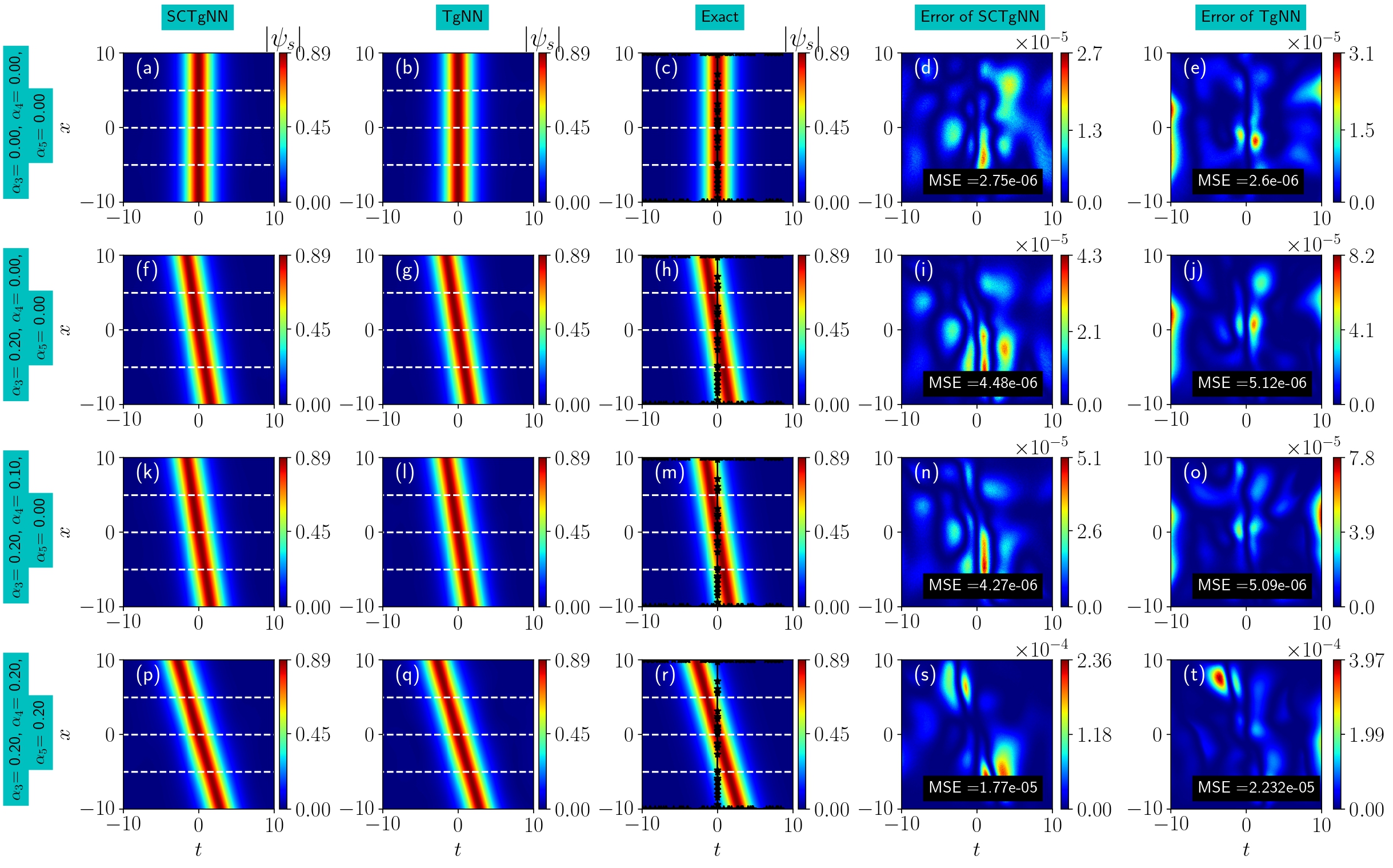}
	\end{center}
	\vspace{-0.3cm}
	\caption{ Result of soliton solution coming out from SCTgNN and TgNN, and their errors with $\alpha_2=1$. The first three columns represent the prediction of SCTgNN, prediction of TgNN and exact result respectively. The error between the prediction of SCTgNN and TgNN are shown in last two columns. The parameters $\alpha_3$, $\alpha_4$ and $\alpha_5$ are displayed in the left side margin.}
	\label{fig2}
\end{figure*}

Figure \ref{fig2} represents the comparison of predicted soliton (using SCTgNN and TgNN models) with analytical soliton for the generalized NLS equation. The first row figures (Fig. \ref{fig2}(a)-\ref{fig2}(c)) are shown for the values, $\alpha_2=1$, $\alpha_3=\alpha_4=\alpha_5=0$ and $c=1$ which is nothing but the solutions of standard NLS equation. Also, the error diagram about the differences between SCTgNN and TgNN is displayed in Figs. \ref{fig2}(d) and \ref{fig2}(e). When we increase the value of $\alpha_3=0.2$ and consider all other parametric values to be the same then the orientation of the soliton slightly changes which is shown in Figs. \ref{fig2}(f)-\ref{fig2}(h). The corresponding error diagrams are shown in Figs. \ref{fig2}(i) and \ref{fig2}(j). Further, adding the value of $\alpha_4$ as $0.1$ and the remaining parameters as considered in the previous case (Figs. \ref{fig2}(f)-\ref{fig2}(h)) then the width of soliton changes (see Figs. \ref{fig2}(k)-\ref{fig2}(m)). For error diagrams see Figs. \ref{fig2}(n) and \ref{fig2}(o). Similarly, when we include the fifth-order dispersion parameter ($\alpha_5$) as $0.2$ and consider all other parameters to remain the same as in the previous case the orientation of the soliton changes which was displayed in Figs. \ref{fig2}(p)-\ref{fig2}(r). The corresponding error diagrams are displayed in Figs. \ref{fig2}(s) and \ref{fig2}(t). The star marker in Figs. \ref{fig2}(c), \ref{fig2}(h), \ref{fig2}(m) and \ref{fig2}(r) denote the randomly chosen data points on the initial and boundary conditions. Here, we consider $2000$ data points for initial and $2000$ data points for boundary conditions including $\alpha_3$, $\alpha_4$ and $\alpha_5$. From this figure, we observe that with the addition of higher-order dispersion parameters $\alpha_3$, $\alpha_4$ and $\alpha_5$ to the standard NLS equation the width and orientation of the soliton changes. Also, we notice that both the SCTgNN and TgNN predicted the exact solutions well, with low mean square error (MSE). The MSEs are also displayed on the error prediction plots.

\begin{figure*}[!ht]
	\begin{center}
		\includegraphics[width=\linewidth]{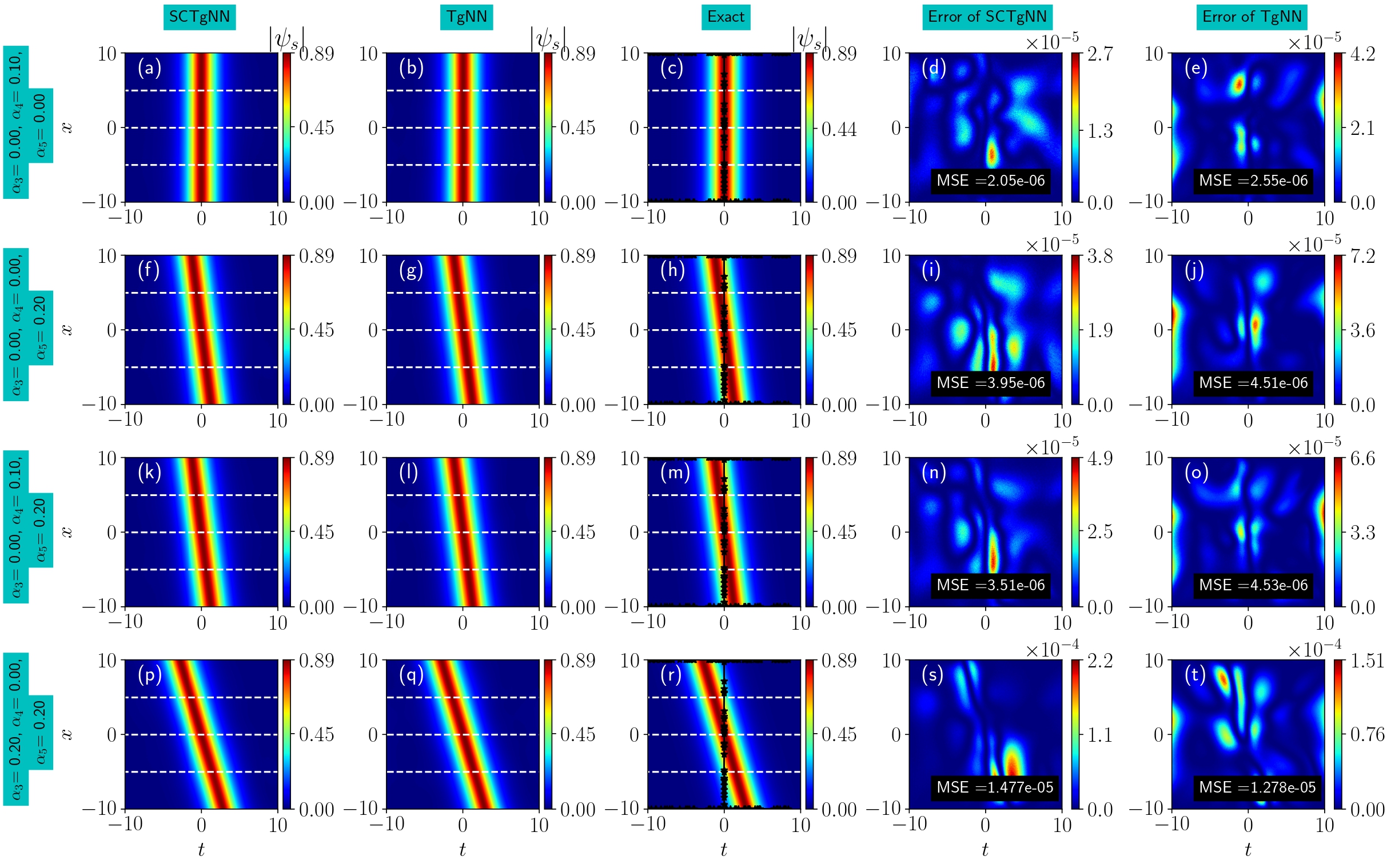}
	\end{center}
	\vspace{-0.3cm}
	\caption{ Result of soliton solution coming out from SCTgNN and TgNN, and their errors with $\alpha_2=1$ for different values of system parameters. The first three columns represent the prediction of SCTgNN, prediction of TgNN and exact result respectively. The error between the prediction of SCTgNN and TgNN are shown in last two columns. The parameters $\alpha_3$, $\alpha_4$ and $\alpha_5$ are displayed in the left side margin.}
	\label{fig21}
\end{figure*}

Figure \ref{fig21} specifically shows the results of the soliton solution of Eq. (\ref{m1am2}) for four different values of system parameters. The first two rows correspond to the LPD (Figs.\ref{fig21}(a)-(e)) and fifth-order NLS (Figs.\ref{fig21}(f)-(j)) equations, respectively, while the other two rows represent randomly chosen system parameters to better understand the behaviour of the system. In the third row (Figs.\ref{fig21}(k)-(o)), for $\alpha_3=0$, $\alpha_4=0.1$ and $\alpha_5=0.2$, the orientation of the soliton is slightly changed from the origin, whereas in the fourth row (Figs.\ref{fig21}(p)-(t)), for different parameters $\alpha_3=0.2$, $\alpha_4=0$ and $\alpha_5=0.2$, the orientation change is large.

From Figs. \ref{fig2} and \ref{fig21}, the MSE for the soliton solution of the Hirota equation is significantly smaller compared to the results of Zhou et al \cite{26}. In their work, the authors predicted soliton solutions with MSE in the range of $10^{-2}$ to $10^{-3}$, while our results show an MSE in the range of $10^{-5}$ to $10^{-6}$. Zhang et al. have proposed an improved physics-informed neural network (IPINN) method to study numerical solutions of the Hirota equation. They predicted soliton solutions and compared their results with PINN. However, they only achieved an error range between $10^{-3}$ and $10^{-4}$ \cite{55}. In contrast, our results showed an error ranging from $10^{-5}$ to $10^{-6}$. From this comparison, it is evident that our SCTgNN model predicts more accurate results.

\begin{table}[!ht]
	\begin{center}
		\begin{tabular}{| m{2cm}| m{2cm} | m{2cm} |m{3cm} |m{3cm}|}
			\hline
			\multicolumn{3}{|c|}{\textbf{Parameters}}& \multicolumn{2}{|c|}{\textbf{MSE}}\\\cline{1-5}
			$\;\;\;\;\;\;\;\alpha_3$  & $\;\;\;\;\;\;\;\alpha_4$  & $\;\;\;\;\;\;\;\alpha_5$ & $\;\;\;\;\;\;$ SCTgNN & $\;\;\;\;\;\;\;$ TgNN\\
			\hline
			$\;\;\;\;\;\;\;$ 0 & $\;\;\;\;\;\;\;$ 0 & $\;\;\;\;\;\;\;$ 0 & $\;\;\;\;\;\;2.73\times 10^{-6}$ & $\;\;\;\;\;\;2.6\times 10^{-6}$\\\hline
			$\;\;\;\;\;\;$ 0.2 & $\;\;\;\;\;\;\;$ 0 & $\;\;\;\;\;\;\;$ 0 & $\;\;\;\;\;\;4.48\times 10^{-6}$ & $\;\;\;\;\;\;5.12\times 10^{-6}$ \\\hline
			$\;\;\;\;\;\;$ 0.2 &  $\;\;\;\;\;\;$ 0.1 & $\;\;\;\;\;\;\;$ 0 & $\;\;\;\;\;\;4.27\times 10^{-6}$ & $\;\;\;\;\;\;5.09\times 10^{-6}$ \\\hline
			$\;\;\;\;\;\;$ 0.2 & $\;\;\;\;\;\;$ 0.2 & $\;\;\;\;\;\;$ 0.2 & $\;\;\;\;\;\;1.77\times 10^{-5}$ & $\;\;\;\;\;\;2.232\times 10^{-5}$ \\\hline
			$\;\;\;\;\;\;$ 0 & $\;\;\;\;\;\;$ 0.1 & $\;\;\;\;\;\;$ 0 & $\;\;\;\;\;\;2.05\times 10^{-6}$ & $\;\;\;\;\;\;2.55\times 10^{-6}$ \\\hline
			$\;\;\;\;\;\;$ 0 & $\;\;\;\;\;\;$ 0 & $\;\;\;\;\;\;$ 0 .2& $\;\;\;\;\;\;3.95\times 10^{-6}$ & $\;\;\;\;\;\;4.51\times 10^{-6}$ \\\hline
			$\;\;\;\;\;\;$ 0 & $\;\;\;\;\;\;$ 0.1 & $\;\;\;\;\;\;$ 0.2 & $\;\;\;\;\;\;3.51\times 10^{-6}$ & $\;\;\;\;\;\;4.53\times 10^{-6}$ \\\hline
			$\;\;\;\;\;\;$ 0.2 & $\;\;\;\;\;\;$ 0 & $\;\;\;\;\;\;$ 0.2 & $\;\;\;\;\;\;1.477\times 10^{-5}$ & $\;\;\;\;\;\;1.278\times 10^{-5}$ \\\hline
		\end{tabular}
		\caption{\label{table1} MSE values for soliton solution using SCTgNN and TgNN.}
	\end{center}
\end{table}

Table \ref{table1} is based on MSE calculations for the soliton solution (the values of first four rows are for Fig.\ref{fig2} and the last four rows are for Fig.\ref{fig21}). It is evident that using data from parameter points in neural network training reveals a low MSE in both SCTgNN and TgNN, as shown in Table \ref{table1}. In simpler terms, SCTgNN and TgNN consistently exhibit smaller MSEs. These findings once again quantitatively demonstrate that SCTgNN and TgNN display superior performance and reliability in investigating solitons in physical systems.

\subsection{Rogue wave solution}
The rogue wave solution for the generalized NLS equation (\ref{m1am2}) is \cite{15}
\begin{subequations}
	\label{m14}
	\begin{align}
	\psi_r=c\left(4\frac{1+2 i B_r x}{D(x,t)}-1\right)e^{i\phi_r x},
	\label{m141}
	\end{align}
	where the function $D(x,t)$ is given by
	\begin{align}
	D(x,t)=1+4B_r^2x^2+4(ct+v_rx)^2,
	\label{m142}
	\end{align}
	in which $B_r$, $\phi_r$ and $v_r$ takes the following form
	\begin{align}
	B_r=2c^2(\alpha_2+6c^2\alpha_4),\quad
	\phi_r=2c^2(\alpha_2+3c^2\alpha_4),\quad
	v_r=2c^3(3\alpha_3+15c^2\alpha_5).\label{m143}
	\end{align}
\end{subequations}
If we consider $t=0$, the solution (\ref{m14}) turns out to be the same with $D(x,t)$ 
\begin{align}
D(x,t)=1+4B_r^2x^2+4v_r^2x^2.
\label{m144}
\end{align}
\begin{figure*}[!ht]
	\begin{center}
		\includegraphics[width=\linewidth]{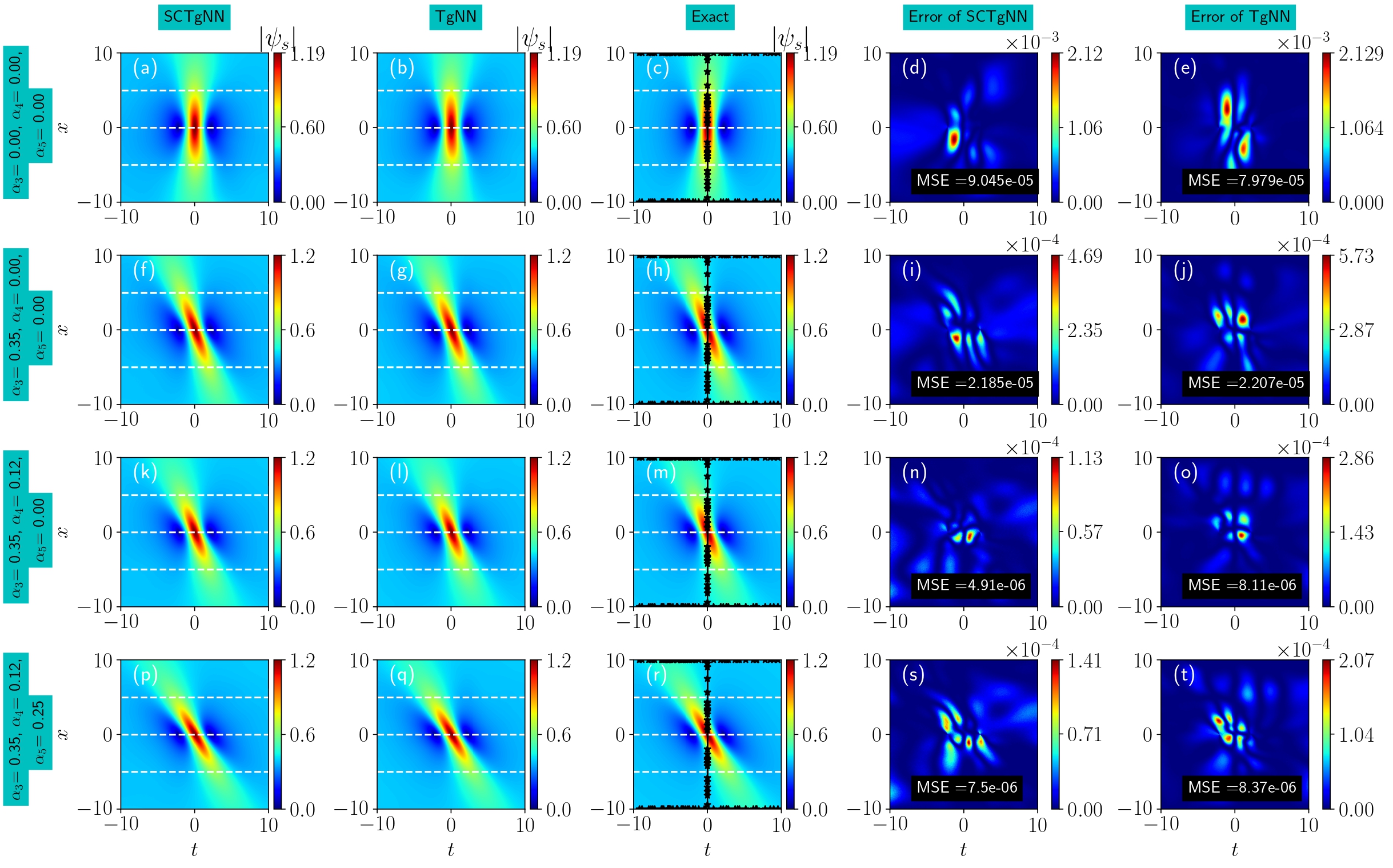}
	\end{center}
	\vspace{-0.3cm}
	\caption{ Result of rogue wave solution coming out from SCTgNN and TgNN, and their errors with $\alpha_2=1$. The first three columns represent the prediction of SCTgNN, prediction of TgNN and exact result respectively. The error between the prediction of SCTgNN and TgNN are shown in the last two columns. The parameters $\alpha_3$, $\alpha_4$ and $\alpha_5$ are displayed in the left side margin.}
	\label{fig3}
\end{figure*}

The prediction of rogue wave solutions and their error estimations obtained from SCTgNN and TgNN are displayed in Figure \ref{fig3} for $\alpha_2=1$ and $c=1$. The first row of figures (Fig. \ref{fig3}(a)-\ref{fig3}(c)) show the solutions of the standard NLS equation with $\alpha_3=\alpha_4=\alpha_5=0$. Additionally, the error diagram showing the differences between SCTgNN and TgNN is displayed in Figs. \ref{fig3}(d) and \ref{fig3}(e). When we increase the value of $\alpha_3$ to $0.2$, while keeping all other parameters the same, we observe a slight change in the orientation of the rogue wave, as shown in Figs. \ref{fig3}(f)-\ref{fig3}(h). The corresponding error diagrams are shown in Figs. \ref{fig3}(i) and \ref{fig3}(j). Next, we add the value of $\alpha_4=0.1$, while maintaining the previous parameter values (Figs. \ref{fig3}(f)-\ref{fig3}(h)), and observe changes in the width of the rogue wave, as seen in Figs. \ref{fig3}(k)-\ref{fig3}(m). The error diagrams are shown in Figs. \ref{fig3}(n) and \ref{fig3}(o). Similarly, when we include the fifth-order dispersion parameter ($\alpha_5=0.2$), while keeping all the other parameters the same as in the previous case, we observe changes in the orientation of the rogue waves, displayed in Figs. \ref{fig3}(p)-\ref{fig3}(r). The corresponding error diagrams are displayed in Figs. \ref{fig3}(s) and \ref{fig3}(t). The star marker in the third column (Figs. \ref{fig3}(c), \ref{fig3}(h), \ref{fig3}(m) and \ref{fig3}(r)) denotes randomly chosen data points on the initial and boundary conditions. We considered $2000$ data points for the initial conditions and $2000$ data points for the boundary conditions, including $\alpha_3$, $\alpha_4$, and $\alpha_5$. From this figure, we observe that with the addition of higher-order dispersion parameters $\alpha_3$, $\alpha_4$, and $\alpha_5$ to the standard NLS equation, the width and orientation of the rogue wave change. In the case of rogue waves also, the SCTgNN and TgNN predictions match well with the exact solution, exhibiting low MSE. The MSE of the prediction is displayed at the bottom of the error prediction plots. These results indicate that the new SCTgNN and TgNN models perform well in predicting the rogue wave solution for the generalized NLS equation, with only a few small MSE errors.

\begin{figure*}[!ht]
	\begin{center}
		\includegraphics[width=\linewidth]{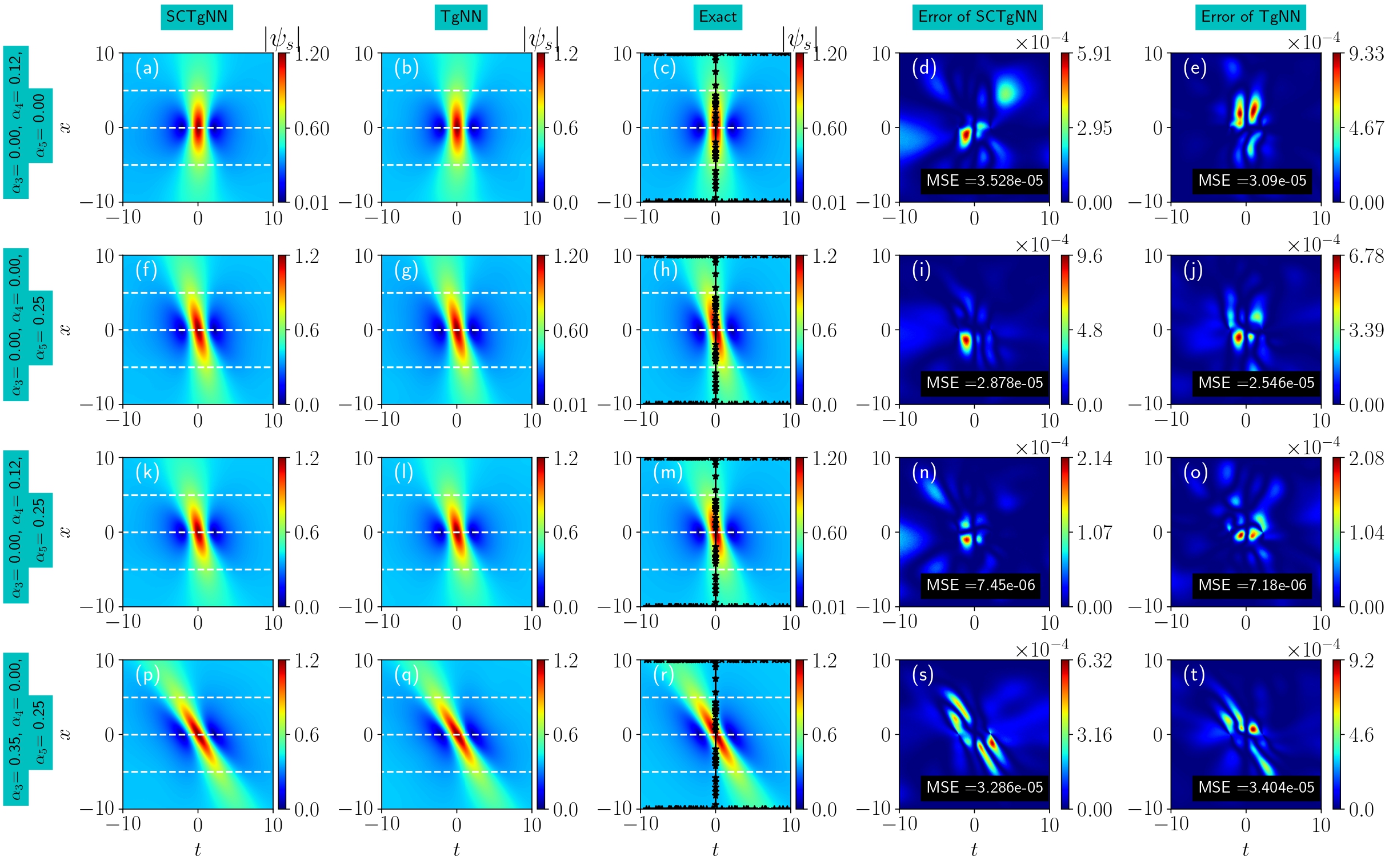}
	\end{center}
	\vspace{-0.3cm}
	\caption{ Result of rogue wave solution coming out from SCTgNN and TgNN, and their errors with $\alpha_2=1$ for different values of system parameters. The first three columns represent the prediction of SCTgNN, prediction of TgNN and exact result respectively. The error between the prediction of SCTgNN and TgNN are shown in the last two columns. The parameters $\alpha_3$, $\alpha_4$ and $\alpha_5$ are displayed in the left side margin.}
	\label{fig31}
\end{figure*}

Figure \ref{fig31} presents the results of the rogue wave solution for the generalized NLS equation (\ref{m1am2}) under four different sets of system parameters. The first two rows correspond to the LPD equations (Figs. \ref{fig31}(a)-(e)) and the fifth-order NLS equations (Figs. \ref{fig31}(f)-(j)), respectively. The remaining two rows illustrate the behaviour of the system under randomly selected parameters. In the third row (Figs. \ref{fig31}(k)-(o)), with parameters $\alpha_3=0$, $\alpha_4=0.12$, and $\alpha_5=0.25$, the orientation of the rogue waves shows a slight deviation from the origin. In the fourth row (Figs. \ref{fig31}(p)-(t)), with different values of $\alpha_3=0.35$, $\alpha_4=0$, and $\alpha_5=0.25$, the orientation change is more pronounced. Additionally, these figures indicate that the width of the rogue waves also decreases.

From Figs. \ref{fig3} and \ref{fig31}, it is evident that the MSE for the rogue wave solution of the Hirota equation is significantly smaller in our results compared to those of Zhou et al \cite{26}. In their work, the authors predicted rogue wave solutions with an MSE in the range of $10^{-2}$, while our results show an MSE in the range of $10^{-5}$ to $10^{-6}$. Zhang et al. also investigated rogue wave solutions of the Hirota equation, similar to their exploration of soliton solutions \cite{55}. However, their error range was larger compared to our results. In Ref. \cite{27}, the MSE for the rogue wave solution for the LPD equation is obtained in the range of $10^{-3}$, whereas in Figs. \ref{fig3}(k)-(o), we obtained the MSE in the range of $10^{-6}$ (a lower MSE indicates higher accuracy). This comparison clearly demonstrates that our SCTgNN model produces more accurate predictions.

\begin{table}[!ht]
	\begin{center}
		\begin{tabular}{| m{2cm}| m{2cm} | m{2cm} |m{3cm} |m{3cm}|}
			\hline
			\multicolumn{3}{|c|}{\textbf{Parameters}}& \multicolumn{2}{|c|}{\textbf{MSE}}\\\cline{1-5}
			$\;\;\;\;\;\;\;\alpha_3$  & $\;\;\;\;\;\;\;\alpha_4$  & $\;\;\;\;\;\;\;\alpha_5$ & $\;\;\;\;\;\;$ SCTgNN & $\;\;\;\;\;\;\;$ TgNN\\
			\hline
			$\;\;\;\;\;\;\;$  0 & $\;\;\;\;\;\;\;$ 0 & $\;\;\;\;\;\;\;$ 0 & $ \;\;\;\;\;\;9.045\times 10^{-5}$ & $\;\;\;\;\;\; 7.979\times 10^{-5}$\\\hline
			$\;\;\;\;\;\;$ 0.35 & $\;\;\;\;\;\;\;$ 0 & $\;\;\;\;\;\;\;$ 0 & $\;\;\;\;\;\; 2.183\times 10^{-5}$ & $\;\;\;\;\;\;2.207\times 10^{-5}$ \\\hline
			$\;\;\;\;\;\;$ 0.35 & $\;\;\;\;\;\;$ 0.12 & $\;\;\;\;\;\;\;$ 0 & $\;\;\;\;\;\;4.91\times 10^{-6}$ & $\;\;\;\;\;\;8.11\times 10^{-6}$ \\\hline
			$\;\;\;\;\;\;$ 0.35 & $\;\;\;\;\;\;$ 0.12 & $\;\;\;\;\;\;$ 0.25 & $\;\;\;\;\;\;7.5\times 10^{-6}$ & $\;\;\;\;\;\;8.37\times 10^{-6}$ \\\hline
			$\;\;\;\;\;\;$ 0 & $\;\;\;\;\;\;$ 0.12 & $\;\;\;\;\;\;$ 0 & $\;\;\;\;\;\;3.5\times 10^{-5}$ & $\;\;\;\;\;\;3.09\times 10^{-5}$ \\\hline
			$\;\;\;\;\;\;$ 0 & $\;\;\;\;\;\;$ 0 & $\;\;\;\;\;\;$ 0.25 & $\;\;\;\;\;\;2.878\times 10^{-5}$ & $\;\;\;\;\;\;2.546\times 10^{-5}$ \\\hline
			$\;\;\;\;\;\;$ 0 & $\;\;\;\;\;\;$ 0.12 & $\;\;\;\;\;\;$ 0.25 & $\;\;\;\;\;\;7.45\times 10^{-6}$ & $\;\;\;\;\;\;7.18\times 10^{-6}$ \\\hline
			$\;\;\;\;\;\;$ 0.35 & $\;\;\;\;\;\;$ 0 & $\;\;\;\;\;\;$ 0.25 & $\;\;\;\;\;\;3.286\times 10^{-5}$ & $\;\;\;\;\;\;3.404\times 10^{-5}$ \\\hline
		\end{tabular}
		\caption{\label{table2} MSE values for the rogue wave solution using SCTgNN and TgNN.}
	\end{center}
\end{table}

Table \ref{table2} presents the results of MSE calculations for the rogue wave solution (the values for the first four rows pertain to Fig. \ref{fig3}, while the values for the last four rows correspond to Fig. \ref{fig31}). It becomes apparent that when employing data from parameter points during neural network training, there is a noticeable lower MSE in both SCTgNN and TgNN, as illustrated in Table \ref{table2}. More precisely, SCTgNN and TgNN consistently display a lower MSE. These results, once more, provide quantitative evidence of SCTgNN's and TgNN's better performance and their trustworthiness in the examination of rogue waves in physical systems.

\subsection{Breather solution}
The breather solution for the Eq. (\ref{m1am2}) is given by \cite{15}
\begin{subequations}
	\label{m15am015}
	\begin{align}
	\psi_b=c\left(1+\frac{\kappa^2 C(x)+i\kappa\sqrt{4-\kappa^2}S(x)}{\sqrt{4-\kappa^2}cos[\kappa(ct+v_bx)]-2C(x)}\right)e^{i\phi_b x}.
	\label{m15}
	\end{align}
	In Eq. (\ref{m15}), the terms $C(x)$ and $S(x)$ have the following form
	\begin{align}
	C(x)=cosh\left(B_b\kappa\sqrt{1-\frac{\kappa^2}{4}x}\right),\quad
	S(x)=sinh\left(B_b\kappa\sqrt{1-\frac{\kappa^2}{4}x}\right).
	\label{m015}
	\end{align}
	The terms $v_b$, $\phi_b$ and $B_b$ in Eq. (\ref{m15}) are denoted by
	\begin{align}
	v_b=&\alpha_3c^3(6-\kappa^2)+\alpha_5c^5(30-10\kappa^2+\kappa^4),\nonumber\\
	\phi_b=&2[\alpha_2c^2+\alpha_4c^4(6-\kappa^2)],\quad
	B_b=2(\alpha_2c^2+3\alpha_4c^4).
	\label{m0151}
	\end{align}
\end{subequations}
If $x=0$, the breather solution takes the form
\begin{align}
\psi_b=c\left(1+\frac{\kappa^2}{\sqrt{4-\kappa^2}cos[\kappa(ct)]-2}\right),
\label{m16}
\end{align} 
where $C(x)$ and $S(x)$ are given in Eq. (\ref{m015}).

\begin{figure*}[!ht]
	\begin{center}
		\includegraphics[width=\linewidth]{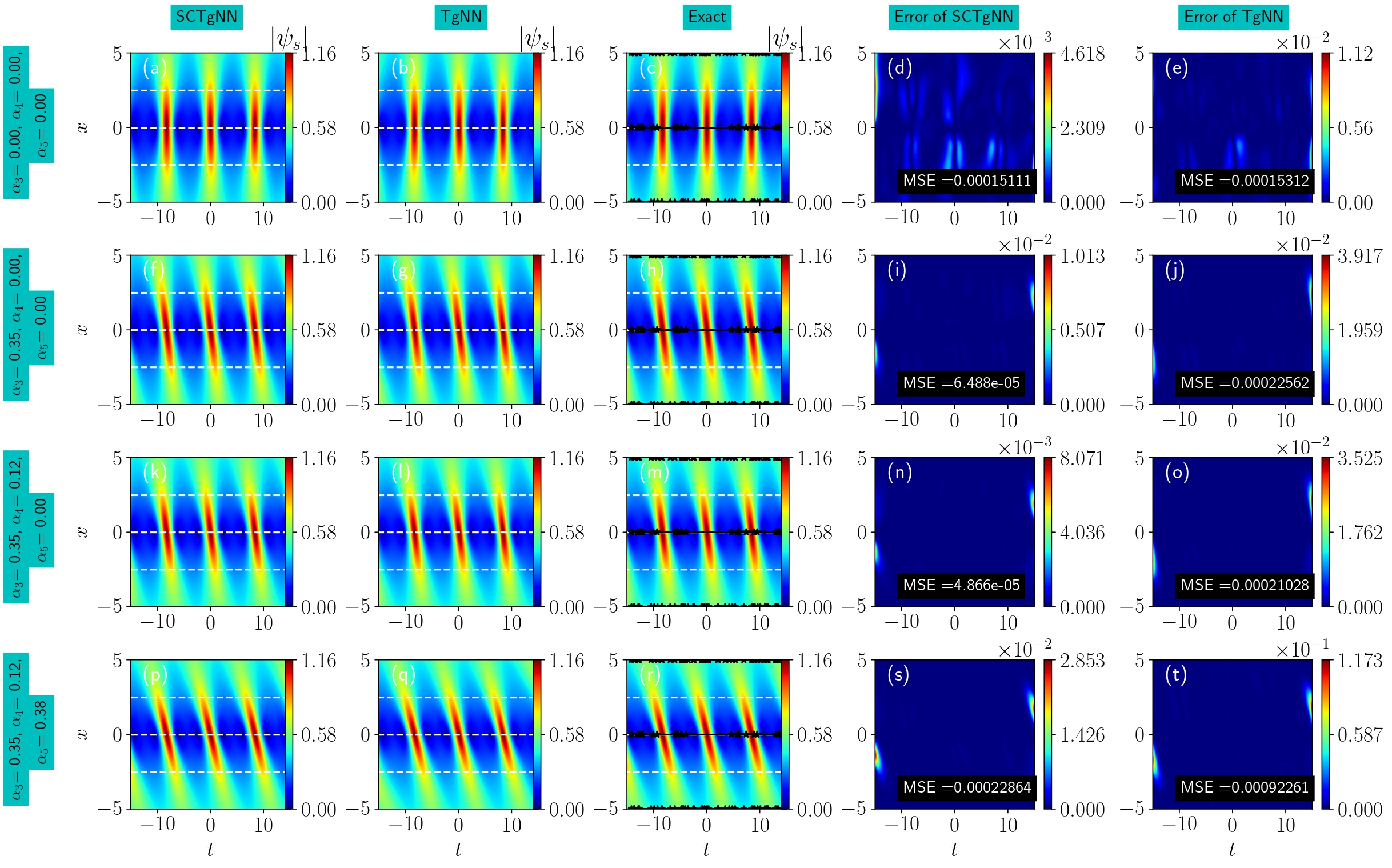}
	\end{center}
	\vspace{-0.3cm}
	\caption{ Result of breather solution coming out from SCTgNN and TgNN, and their errors with $\alpha_2=1$. The first three columns are represent the prediction of SCTgNN, prediction of TgNN and exact result respectively. The error between the prediction of SCTgNN and TgNN are shown in last two columns. The parameters $\alpha_3$, $\alpha_4$ and $\alpha_5$ are displayed in the left side margin.}
	\label{fig4}
\end{figure*}

In Fig. \ref{fig4}, we present the results of the dynamical prediction of breather solutions and their error estimations using SCTgNN and TgNN, where $\alpha_2=1$ and $c=1$. Unlike solitons and rogue waves, breathers can be predicted using the following initial values when $x=0$. Here, $t$ ranges from $-15$ to $15$ and $x$ ranges from $-5$ to $5$, while the other $\alpha_i$'s remain the same. The figures in the first row (Fig. \ref{fig4}(a)-\ref{fig4}(c)) represent the solution of the standard NLS equation, corresponding to $\alpha_3=\alpha_4=\alpha_5=0$. Additionally, Figs. \ref{fig4}(d) and \ref{fig4}(e) display the error diagram, highlighting the discrepancies between SCTgNN and TgNN. Upon increasing the value of $\alpha_3$ to $0.2$, while maintaining the other parameters constant, we observe subtle changes in the orientation of the breathers, as depicted in Figs. \ref{fig4}(f)-\ref{fig4}(h). The corresponding error diagrams are provided in Figs. \ref{fig4}(i) and \ref{fig4}(j). Furthermore, when we set $\alpha_4=0.1$ and keep the parameters consistent with the previous case (Figs. \ref{fig4}(f)-\ref{fig4}(h)), the width of the breathers undergoes alterations, as shown in Figs. \ref{fig4}(k)-\ref{fig4}(m). The error diagrams for this situation can be observed in Figs. \ref{fig4}(n) and \ref{fig4}(o). Similarly, by introducing the fifth-order dispersion parameter ($\alpha_5=0.2$) while maintaining the other parameters from the previous case, we again observe changes in the orientation of the breathers, as illustrated in Figs. \ref{fig4}(p)-\ref{fig4}(r). The corresponding error diagrams are presented in Figs. \ref{fig4}(s) and \ref{fig4}(t). In Figs. \ref{fig4}(c), \ref{fig4}(h), \ref{fig4}(m), and \ref{fig4}(r), the star marker indicates randomly chosen data points on the initial and boundary conditions. For our analysis, we used $2000$ data points for each condition, including $\alpha_3$, $\alpha_4$, and $\alpha_5$.
Overall, the results demonstrate that the inclusion of higher-order dispersion parameters $\alpha_3$, $\alpha_4$, and $\alpha_5$ in the standard NLS equation leads to changes in the width and orientation of the breathers. 
Similarly, the SCTgNN and TgNN models demonstrate superior error prediction capabilities, as illustrated in the MSE prediction figures. These findings indicate that the new SCTgNN and TgNN models excel at estimating the breather solution for the generalized NLS equation, with only a few minor errors.

\begin{figure*}[!ht]
	\begin{center}
		\includegraphics[width=\linewidth]{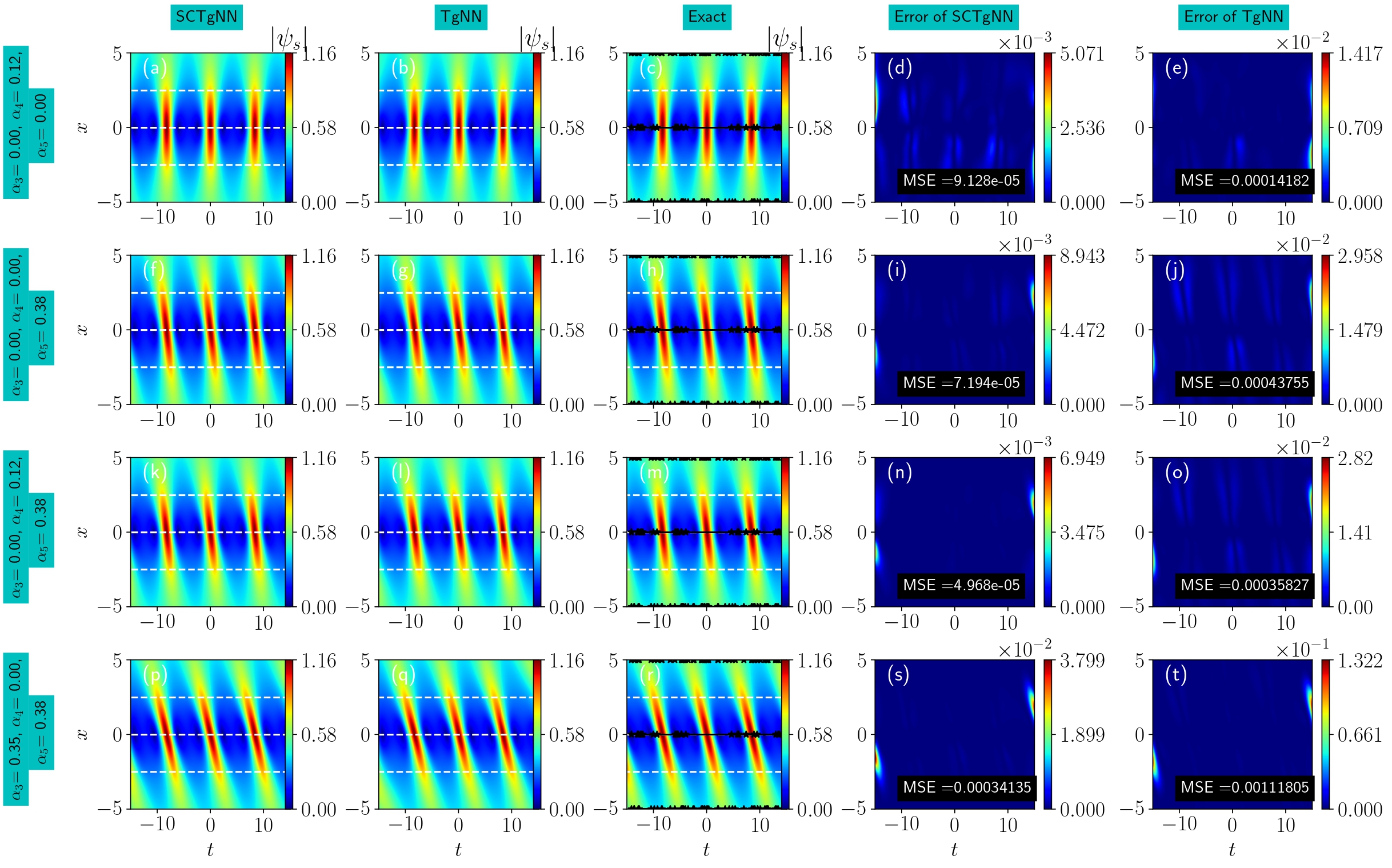}
	\end{center}
	\vspace{-0.3cm}
	\caption{ Result of breather solution coming out from SCTgNN and TgNN, and their errors with $\alpha_2=1$ for different values of system parameters. The first three columns are represent the prediction of SCTgNN, prediction of TgNN and exact result respectively. The error between the prediction of SCTgNN and TgNN are shown in last two columns. The parameters $\alpha_3$, $\alpha_4$ and $\alpha_5$ are displayed in the left side margin.}
	\label{fig41}
\end{figure*}

Figure \ref{fig41} shows the results of breathers for the generalized NLS equation (\ref{m1am2}) for four different sets of system parameters. The first two rows depict the LPD equations (Figs. \ref{fig41}(a)-(e)) and the fifth-order NLS equations (Figs. \ref{fig41}(f)-(j)). The last two rows represent the system's behaviour with randomly chosen parameters. In the third row (Figs. \ref{fig41}(k)-(o)), with parameters $\alpha_3=0$, $\alpha_4=0.12$, and $\alpha_5=0.38$, the orientation of the breathers shows a slight deviation from the origin. In the fourth row (Figs. \ref{fig41}(p)-(t)), with different values of $\alpha_3=0.35$, $\alpha_4=0$, and $\alpha_5=0.38$, the orientation change is more pronounced.

Similarly, as shown in Figs. \ref{fig4} and \ref{fig41}, our results for the MSE of the breather solution of the Hirota equation are significantly smaller compared to those of Zhou et al \cite{26}. While their work reported breather solutions with an MSE of $10^{-2}$, our results exhibit an MSE ranging from $10^{-3}$ to $10^{-5}$. This indicates that our SCTgNN model provides much more accurate predictions.

\begin{table}[!ht]
	\begin{center}
		\begin{tabular}{| m{2cm}| m{2cm} | m{2cm} |m{3cm} |m{3cm}|}
			\hline
			\multicolumn{3}{|c|}{\textbf{Parameters}}& \multicolumn{2}{|c|}{\textbf{MSE}}\\\cline{1-5}
			$\;\;\;\;\;\;\;\alpha_3$  & $\;\;\;\;\;\;\;\alpha_4$  & $\;\;\;\;\;\;\;\alpha_5$ & $\;\;\;\;\;\;$ SCTgNN & $\;\;\;\;\;\;\;\;$TgNN\\
			\hline
			$\;\;\;\;\;\;\;$ 0 & $\;\;\;\;\;\;\;$ 0 & $\;\;\;\;\;\;\;$ 0 & $\;\;\;\;\;\;1.51\times 10^{-4}$ & $\;\;\;\;\;\;1.33\times 10^{-4}$\\\hline
			$\;\;\;\;\;\;$ 0.35 & $\;\;\;\;\;\;\;$ 0 & $\;\;\;\;\;\;\;$ 0 & $\;\;\;\;\;\;6.50\times 10^{-5}$ & $\;\;\;\;\;\;2.26\times 10^{-4}$ \\\hline
			$\;\;\;\;\;\;$ 0.35 & $\;\;\;\;\;\;$ 0.12 & $\;\;\;\;\;\;\;$ 0 & $\;\;\;\;\;\;4.90\times 10^{-5}$ & $\;\;\;\;\;\;2.10\times 10^{-4}$ \\\hline
			$\;\;\;\;\;\;$ 0.35 & $\;\;\;\;\;\;$ 0.12 & $\;\;\;\;\;\;$ 0.38 & $\;\;\;\;\;\;2.29\times 10^{-4}$ & $\;\;\;\;\;\;9.23\times 10^{-4}$ \\\hline
			$\;\;\;\;\;\;$ 0 & $\;\;\;\;\;\;$ 0.12 & $\;\;\;\;\;\;$ 0 & $\;\;\;\;\;\;9.128\times 10^{-5}$ & $\;\;\;\;\;\;1.4182\times 10^{-4}$ \\\hline
			$\;\;\;\;\;\;$ 0 & $\;\;\;\;\;\;$ 0 & $\;\;\;\;\;\;$ 0.38 & $\;\;\;\;\;\;7.194\times 10^{-5}$ & $\;\;\;\;\;\;4.3755\times 10^{-4}$ \\\hline
			$\;\;\;\;\;\;$ 0 & $\;\;\;\;\;\;$ 0.12 & $\;\;\;\;\;\;$ 0.38 & $\;\;\;\;\;\;4.96\times 10^{-5}$ & $\;\;\;\;\;\;3.5827\times 10^{-4}$ \\\hline
			$\;\;\;\;\;\;$ 0.35 & $\;\;\;\;\;\;$ 0 & $\;\;\;\;\;\;$ 0.38 & $\;\;\;\;\;\;3.41\times 10^{-4}$ & $\;\;\;\;\;\;1.11\times 10^{-3}$ \\\hline
		\end{tabular}
		\caption{\label{table3} MSE values for the breather solution using SCTgNN and TgNN.}
	\end{center}
\end{table}

Table \ref{table3} showcases the outcomes of MSE computations pertaining to the breather solution (the values of the initial four rows relate to Fig. \ref{fig4}, whereas the subsequent four rows pertain to Fig. \ref{fig41}). It becomes evident that when incorporating data from parameter points into the neural network training process, there is a conspicuous low MSE in SCTgNN and TgNN, as depicted in Table \ref{table3}. In simpler terms, SCTgNN and TgNN consistently demonstrate a lower MSE. These findings reaffirm that SCTgNN and TgNN exhibit better performance and reliability in investigating breathers in physical systems, substantiating their quantitative excellence once again.
\section{Conclusion}
In our study, we have explored the generalized NLS equation and their localized solutions using a novel network structure known as SCTgNN, which incorporates the concepts of both the PINN and TgNN models. Under specific conditions, this generalized equation augments results of the NLS equation, Hirota equation, LPD equation and the fifth-order equation.  Hence, the detailed study on this equation benefits physical community, in particular the optics and hydrodynamics community. By synergizing the strengths of TgNN and SCPINN, our aim was to enrich our comprehension of intricate phenomena, particularly in scenarios where standardized patterns do not apply, such as in nonlinear systems. Furthermore, our scope extended beyond the exclusive prediction of rogue wave patterns, as observed in a previous study \cite{45} employing the TgNN model. Instead, we broadened our investigation to encompass a spectrum of wave behaviours, including solitons, rogue waves, and breathers, within the broader framework of the generalized NLS equation. Additionally, we successfully predicted solitons, rogue waves and breathers for the generalized NLS equation using our new SCTgNN model. To evaluate the efficacy of our innovative SCTgNN model and the established TgNN model, we conducted comprehensive MSE predictions. These MSE predictions are well-matched with exact values, yielding very low errors in both models. This approach enhances our capacity to predict and comprehend intricate systems characterized by complex behaviours.

Our findings confirm that adding higher-order dispersion parameters to the standard NLS equation leads to changes in the width and orientation of solitons, rogue waves, and breathers. Interestingly, both SCTgNN and TgNN models consistently predict errors with high accuracy. We have integrated system parameters $\alpha_2$, $\alpha_3$, $\alpha_4$, and $\alpha_5$ into the initial conditions. This provides a significant advantage: the straightforward prediction of solutions for various parameter values. Furthermore, this approach extends beyond the specified range, enabling predictions for a large number of value. This expansion enhances both the applicability and the value of our work. These results highlight the strength of the new SCTgNN model and established TgNN model in closely approximating soliton, rogue waves, and breather solutions for the generalized NLS equation, even as we consider more complex scenarios with higher-order effects. Recently, Ablowitz and his collaborators introduced the fractional integrable system for the first time \cite{56}. They derived the fractional nonlinear Schr\"{o}dinger (fNLS) and fractional Korteweg-de Vries (fKdV) equation, and their corresponding one-soliton solutions, by employing the inverse scattering transform (IST) method. In near future, we intend to extend our present analysis to fractional types of integrable equations. We also plan to study the higher-order solutions for the considered system.
\section{Acknowledgments}
KT thanks Science and Engineering Research Board, Government of India, under the Grant No. CRG/2021/002428. NS wishes to thank DST-SERB, Government of India for providing National Post-Doctoral Fellowship under Grant No. PDF/2023/000619. NVP wishes to thank Department of Science and Technology (DST), India, for the financial support under Women Scientist Scheme-A. The work of M.S. was supported by the Science and Engineering Research Board, Government of India, under the Grant No. CRG/2021/002428. MS also acknowledges MoE-RUSA 2.0 Physical Sciences, Government of India for sponsoring this research work.

\end{document}